\documentclass[secnumarabic,amssymb, amsmath,showpacs,  
nofootinbib,tightenlines, nobibnotes, aps, prl]{revtex4}
\usepackage{graphicx}
\begin{document}
\newcommand{\bbc}{\begin{center}}
\newcommand{\eec}{\end{center}}
\newcommand{\be}{\begin{equation}}
\newcommand{\ee}{\end{equation}}
\newcommand{\ba}{\begin{eqnarray}}
\newcommand{\ea}{\end{eqnarray}}

\bigskip
\vspace{2cm}
\title{Radiative two-pion decay of the tau lepton}
\vskip 6ex
\author{A. Flores-Tlalpa}
\email{afflores@fis.cinvestav.mx}
\author{G. L\'opez Castro}
\email{glopez@fis.cinvestav.mx}
\affiliation{Departamento de F\'{\i}sica, Cinvestav,  \\
Apartado Postal 14-740, 07000 M\'exico, D.F. M\'exico}
\author{G. Toledo S\'anchez}
\email{toledo@fisica.unam.mx}
\affiliation{Instituto de F\'\i sica, UNAM, Apdo. Postal 20-364 \\
01000 M\'exico, D.F. M\'exico}
\bigskip

\bigskip

\bigskip

\begin{abstract}
We consider the bremsstrahlung and model-dependent contributions to the 
radiative decay $\tau^- \to \pi^-\pi^0\nu_{\tau}\gamma$ in the context of 
a meson dominance model. We focus on several observables 
related 
to this decay, including the branching ratio and the photon and di-pion  
spectra. Particular attention is paid to the sensitivity of 
different observables upon the effects of model-dependent 
contributions and of the magnetic dipole moment of the $\rho^-(770)$ 
vector meson. Important numerical differences are found  with 
respect to results obtained in the framework of chiral perturbation 
theory. 
 \end{abstract}


\maketitle
\bigskip

\bigskip

\section{1. Introduction}

   Semileptonic tau lepton decays are a rich source of information 
about the properties of hadronic resonances below the tau lepton mass 
scale. 
They provide a clean  environment to study the properties of charged 
$\rho(770)$ and $a_1(1260)$ resonances which otherwise would be produced 
only through purely hadronic processes. The interplay of strong, weak 
and electromagnetic interactions in such processes offers an interesting 
place to test models for these interactions at low energies and to extract 
information about fundamental parameters of the standard model 
\cite{Davier05}.  

In this paper we are interested in the study of the 
radiative $\tau^- \to \pi^-\pi^0\nu_{\tau}\gamma$ decay, a process that 
involves simultaneously the three fundamental interactions at the lowest 
order. This 
decay channel has been studied previously in references \cite{genaro00, 
Cirigliano02} within different models and with different purposes. As is 
well 
known, the corresponding non-radiative $\tau^- \to \pi^-\pi^0\nu_{\tau}$ 
decay is dominated by the  production of the $\rho^-(770)$ vector meson; 
thus, the emission of a single photon from this process is expected to 
carry  information about the $\rho^-$-meson magnetic dipole moment 
\cite{genaro00}. A   
meaningful extraction of this property from data is possible only with a 
full account of the model-dependent contributions to the radiative decay, 
which was not included in ref. \cite{genaro00}. In this paper we 
pursue this study and consider 
the complete calculation of the radiative amplitude using a 
phenomenological 
model that includes all possible intermediate hadronic states. 

A different approach is followed in ref. \cite{Cirigliano02}, where
the radiative amplitudes were calculated in the framework of chiral 
perturbation theory and including resonances in 
the relevant kinematical regions. The interest of Ref. 
\cite{Cirigliano02} was focused on the 
relationship between the di-pion tau decay data and the its leading 
hadronic contribution to the anomalous magnetic  moment of the muon 
$a_{\mu}$ \cite{davier98}.
 As is known, present experimental information on 
$\tau \to \pi\pi\nu$ decays are photon inclusive measurements 
\cite{Davier05}. Thus, removing  radiative effects from the 
measured di-pion mass distribution in such decays is  
important to predict the leading order hadronic vacuum 
polarization contribution to $a_{\mu}$. A comparison of the two-pion mode 
in tau decays and $e^+e^-$ annihilations provides a sensitive test of the 
CVC hypothesis. At present, the prediction of  $a_{\mu}^{had}$ based on 
$\tau  \to \pi\pi\nu$ data seems to exceed by more than two 
standard deviations the corresponding prediction based on $e^+e^-$ data 
\cite{Davier05}, even after the known sources of isospin breaking 
corrections are removed \cite{Cirigliano02, Cirigliano01, ghozzi}. Since 
the production of high energy photons in $\tau^-  \to \pi^-\pi^0\nu\gamma$ 
decays is driven by the model-dependent contributions, a good account of 
the model-dependent effects is again mandatory.

   This paper is organized as follows: in section 2 we describe the 
necessary one-loop modifications of the propagator and electromagnetic 
vertex of the unstable $\rho^-$ vector meson to achieve a gauge-invariant 
amplitude for the model-independent contributions; in section 3 we 
describe the form of the amplitude for the non-radiative $\tau$ lepton  
decay and fix the parameters involved in our approximation; in section 4 
we focus on the different contributions to the radiative decay amplitude 
and check their gauge invariance requirements; in section 5 we fix the 
coupling constants involved in the model-independent contributions and 
compute the different observables associated to the radiative two-pion 
$\tau$ lepton decays; our conclusions are summarized in section 6 and an 
Appendix is devoted to discuss the kinematics associated to the this 
four-body decay.

\section{2. Gauge invariance and unstable particles in radiative 
processes}

  The physical amplitudes of radiative processes (${\cal 
M}=\epsilon^{\mu}{\cal M}_{\mu}$, $\epsilon$ being the photon polarization 
four-vector) have to satisfy the 
electromagnetic gauge invariance condition $k^{\mu}{\cal M}_{\mu}=0$, 
where $k$ is the photon four-momentum. As it has been discussed elsewhere 
\cite{unstable}, when charged unstable particles are produced as 
intermediate 
states of a physical process some care must be taken to make 
compatible the unstable character of the resonance with the gauge 
invariance condition. One of the proposals to deal with 
this problem is the so-called {\it fermion loop-scheme} ({\it fls}) 
for gauge bosons \cite{fls}. 
According to the {\it fls}, only the fermion contributions in loop 
corrections 
to the propagator and electromagnetic vertex of gauge bosons 
have to be included to render gauge-invariant the resonant amplitudes 
\cite{fls}. 

  In the case of hadronic resonances such as the $\rho^-$ meson, it has 
been suggested that an analogous {\it  boson loop-scheme} ({\it 
bls}) \cite{genaro00} can be  used to avoid such potential 
gauge pathologies. It 
has been shown \cite{genaro00, fls} that when particles in loop 
corrections 
are massless, the 
corresponding dressed Green functions obtained in the {\it fls} and {\it 
bls } are the same as the ones obtained using the complex-mass 
prescription $M_0^2 \to 
M^2-iM\Gamma$ ($M$ and $\Gamma$ are the mass and decay width of 
the resonance) in 
the bare 
Green functions. This prescription have been  successfully used 
\cite{mariano} to describe experimental data of the elastic and radiative 
$\pi^+ p$ scattering to extract the mass, width and magnetic moment of the 
$\Delta^{++}$ baryon resonance.

   According to the {\it boson loop-scheme}, one has to include the 
absorptive parts of the one-loop corrections to the electromagnetic vertex 
and the propagator of the resonance in order to satisfy 
electromagnetic gauge invariance \cite{genaro00}. In the case of a 
$\rho^-$ vector-meson of mass $m_{\rho}$ and four-momentum $q$, the 
one-loop absorptive corrections arising from $\pi^-\pi^0$ meson 
loops\footnote{The contribution of loops with $K^-K^0$ mesons can be 
included in a similar way, but we neglect its small contribution in this 
paper.} gives the resonant propagator \cite{genaro00}: 
\be
D^{\mu\nu}_{\rho^-}(q)=-i\frac{g^{\mu\nu}-\frac{\textstyle 
q^{\mu}q^{\nu}}{\textstyle m_{\rho}^2} 
\left(1+i\frac{\textstyle \Gamma_{\rho}(q^2)}{\textstyle \sqrt{q^2}} 
\right)}
{q^2-m_{\rho}^2+i\sqrt{q^2}\Gamma_{\rho}(q^2)}\ ,
\ee
where we have defined the energy-dependent  width (in the limit of 
isospin symmetry $m_{\pi^-}=m_{\pi^0}=m_{\pi}$) as follows:
\[
\Gamma_{\rho}(q^2)=\frac{g_{\rho\pi\pi}^2}{48\pi 
q^2}(q^2-4m_{\pi}^2)^{3/2}\theta (s-4m_{\pi}^2)\ , 
\]
with $g_{\rho\pi\pi}$ the $\rho\pi\pi$ coupling constant (its value is 
discussed below). 

The one-loop absorptive corrections to the electromagnetic vertex (using 
the convention  $\rho^{-\alpha}(p) \to 
\rho^{-\beta}(p')\gamma^{\delta}(k)$ 
for Lorentz indices and four-momenta) 
gives the following result \cite{genaro00}:
\be
ie\Gamma^{\alpha\beta\delta} =  
ie(\Gamma_0^{\alpha\beta\delta}+\Gamma_1^{\alpha\beta\delta}),
\ee
where 
\be \Gamma_0^{\alpha\beta\delta} = (p+p')^\alpha g^{\beta\delta} +
(k^\beta g^{\alpha\delta} - k^\delta g^{\alpha\beta})\beta(0) -
p^\beta g^{\alpha\delta} - p'^\delta g^{\alpha\beta}\ ,
\ee
is the electromagnetic vertex at the tree-level, and $\beta(0)$ the 
value of the magnetic dipole moment of the $\rho^-(770)$ meson in units of 
$e/2m_{\rho}$  ($\beta(0)=2$ corresponds 
to the normal or canonical 
value of the magnetic dipole moment; typical values of 
$\beta(0)$ computed in quark models lies in the interval $1.8 \leq 
\beta(0) 
\leq 3.0$ \cite{mdm}). The absorptive part of the $\pi^-\pi^0$ one-loop 
correction to the electromagnetic vertex of the $\rho^-$ has been computed 
in ref \cite{genaro00} using cutting rules. Its explicit form in the limit 
of isospin symmetry   
is given by:
\begin{eqnarray}
\Gamma_1^{\alpha\beta\delta} & = & \frac{i g_{\rho\pi\pi}^2}{16 \pi (p^2 - 
p'^2)}
\bigg\{ A(p^2) p^\alpha T^{\beta\delta}(p) - A(p'^2) p'^\alpha
T^{\beta\delta}(p') + B(p^2) F^{\alpha\beta}(p) k^\delta + B(p'^2)
F^{\alpha\delta}(p') k^\beta \nonumber\\
 & & + \Big[ A(p^2) + B(p^2) \Big] \Big[ F^{\alpha\beta}(p)
F^{\eta\delta}(p) + F^{\alpha\delta}(p) F^{\eta\beta}(p) \Big] p_\eta
\nonumber\\ & & - \Big[ A(p'^2) + B(p'^2) \Big] \Big[
F^{\alpha\beta}(p') F^{\eta\delta}(p') + F^{\alpha\delta}(p')
F^{\eta\beta}(p') \Big] p'_\eta \bigg\}
\end{eqnarray}
where  :
\begin{eqnarray}
B(q) & = & 2m_{\pi}^2 \ln \left| \frac{q^2 +
(q^4-4m_{\pi}^2q^2)^{1/2}}{q^2-(q^4-4m_{\pi}^2q^2)^{1/2}} \right|
- (q^4-4m_{\pi}^2q^2)^{1/2} \nonumber \\ 
A (q^2) & = & \frac{2 (q^4 - 4m_{\pi}^2q^2)^{3/2}}{3 q^4} \nonumber \\   
 F^{\mu\nu}(q) & = & g^{\mu\nu} - \frac{q^{\mu} k^{\nu}}{q \cdot 
k}, \ \ \ T^{\mu\nu}(q)  =  g^{\mu\nu} - \frac{q^{\mu} q^{\nu}}{q^2}\ . 
\end{eqnarray}

Since the above  Green functions satisfy the Ward 
identity \cite{genaro00} $k^{\alpha}\Gamma_{\alpha\beta\delta}= 
[iD_{\beta\delta}(p)]^{-1} -[iD_{\beta\delta}(p')]^{-1}$, the radiative 
amplitudes involving such 
vertices and propagators is automatically gauge-invariant. We 
will use this prescription in computing the radiative amplitude of $\tau^- 
\to \pi^-\pi^0\nu_{\tau}\gamma$ decay; as it will be discussed below, 
the model-independent contribution to this process involves the production 
and decay of an intermediate  $\rho^-(770)$ vector meson.

\section{3. Non-radiative two-pion decay}

  In this section we focus on the meson dominance model for  
the non-radiative  $\tau^-(P) \to \pi^-(Q)\pi^0(Q')\nu_{\tau}(P')$ decay, 
where the particles four-momenta  are indicated within parenthesis. Our 
phenomenological model is based on the quantum-mechanical requirement of 
unitarity, according to which all possible intermediate states that are 
allowed to contribute given their quantum numbers have to be included (see 
Figure 1). In practice, only a 
few low-lying meson states are sufficient to describe experimental data. 
As it can be verified  
below, this model reproduces the  K\"uhn and Santamaria \cite{ks} 
parametrization of the vector form factor which contains the sum of the 
$\rho^-(770)$ and of its higher excitations.

In the limit of the isospin symmetry, the amplitude for this decay can be 
written in terms of a single vector form factor:
\be
{\cal M}_0 =\frac{G_FV_{ud}}{\sqrt{2}}l^{\mu}(Q-Q')_{\mu} f_+(\tilde{t})\ 
,
\ee
where $G_F$ is the Fermi 
coupling constant, $l^{\mu}=\bar{u}(P')\gamma^{\mu}(1-\gamma_5)u(P)$ 
denotes 
the leptonic current, $\tilde{t}=(Q+Q')^2$ is the square of the momentum 
transfer and 
$V_{ud}$ is the Cabibbo-Kobayashi-Maskawa mixing matrix element. 

For the purposes of illustrating how the model works, we 
will assume that the amplitude is dominated by the exchange of two  
intermediate resonances: the $\rho^-(770)$ and the ${\rho'}^-(1450)$ 
vector mesons as shown in Figure 1. Applying the Feynman rules to the 
diagram of Figure 1 and 
using the vector-meson propagator given in  eq. (1) we  can obtain the 
following expression for  the form factor:
\ba
f_+(\tilde{t})&=&\frac{g_{\rho}g_{\rho\pi\pi}}{m_{\rho}^2-\tilde{t} 
-i\sqrt{\tilde{t}} 
\Gamma_{\rho}(\tilde{t})} 
+\frac{g_{\rho'}g_{\rho'\pi\pi}}{m_{\rho'}^2-\tilde{t}-im_{\rho'}
\Gamma_{\rho'}} \nonumber  \\  
&=& 
\frac{\sqrt{2}}{1+\sigma}\left\{ 
\frac{m_{\rho}^2}{m_{\rho}^2-\tilde{t}-i\sqrt{\tilde{t}} 
\Gamma_{\rho}(\tilde{t})}+ \sigma 
\frac{m_{\rho'}^2}{m_{\rho'}^2-\tilde{t}-im_{\rho'}\Gamma_{\rho'}}\ 
\right\},
\ea
where $g_{\rho}\ (g_{\rho'})$ denotes the weak coupling of the $\rho 
(770)\ (\rho'(1450))$ vector meson (we neglect the energy dependence of 
the decay width of the $\rho'$ meson). 

\begin{figure}
  \includegraphics[width=12cm]{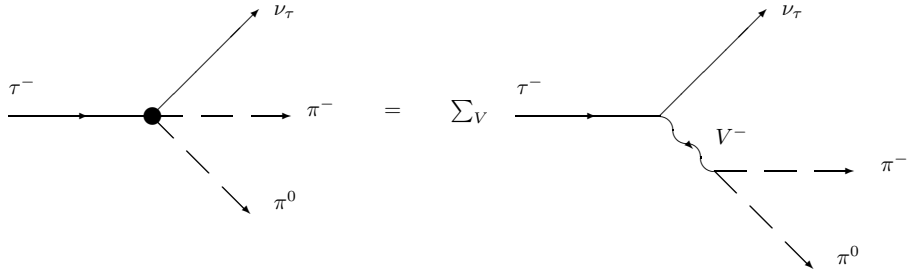}\\
  \caption{Meson dominance model of the non-radiative decay 
$\tau^- \to \pi^-\pi^0\nu_{\tau}$ .}\label{fig1}
\end{figure}

 The expression for the form factor in the second line of eq. (7), which 
coincides with the model of ref. 
\cite{ks}, follows from imposing the normalization 
condition $f_+(\tilde{t}=0)=\sqrt{2}$ and from the definition 
of the parameter $\sigma \equiv 
(m_{\rho}^2g_{\rho'}g_{\rho'\pi\pi})/(m_{\rho'}^2g_{\rho}g_{\rho\pi\pi})$. 
Using the experimental data on the $\rho$ and $\rho'$ decays \cite{pdg} 
(we take  $\Gamma(\rho'\to e^+e^-)=1.48$ keV, $\Gamma(\rho' \to 
\pi^+\pi^-)=26.9$ MeV, which we have estimated from relevant inputs in 
ref. \cite{pdg}), 
we can obtain the following estimate for the relative strengths of these 
vector meson contributions in eq. (7):
\ba \sigma &=& 
\sqrt{\frac{m_{\rho'} \Gamma(\rho' \to 
e^+e^-)\Gamma(\rho'\to \pi\pi)(m_{\rho}^2-4m_{\pi}^2)^{3/2}}{m_{\rho} 
\Gamma(\rho \to e^+e^-)\Gamma(\rho\to 
\pi\pi)(m_{\rho'}^2-4m_{\pi}^2)^{3/2} }} \nonumber \\
&\approx & 0.102\ .
\ea 
This estimate is very close to the experimental value ($|\sigma^{exp}| = 
0.120 \pm 0.008$) reported in ref. \cite{aleph05}(see also our fit 
discussed after eq. 10). This agreement renders 
confidence on the meson dominance model for the radiative decays to be 
discussed in this paper, and in particular about the values of the 
coupling constants extracted from other independent measured processes 
(see section 5).

In order to provide a comparison with the results obtained for radiative 
$\tau$ lepton decays in chiral perturbation theory 
\cite{Cirigliano02}, hereafter we will restrict to the 
model with a single resonance, namely the $\rho(770)$. In order to fix 
the parameters of this model, we have fitted the data of 
ref. \cite{aleph} for the pion form factor below $\sqrt{\tilde{t}}\approx 
1.1$ GeV  and have found that the approximation of using only one 
resonance gives a good description of data with the following 
central values for the resonance parameters:
\be
m_{\rho}= 776.66\ \mbox{\rm MeV},\ \ \ g_{\rho\pi\pi} = 5.488 \ .
\ee
Using these values for the resonance parameters, we obtain the following 
result for the non-radiative branching fraction:
\be
B(\tau^- \to \pi^-\pi^0\nu_{\tau})= 20.75 \%
\ee
Clearly, our simple model underestimates the experimental value whose   
present world average is $B^{exp}(\tau^- \to \pi^-\pi^0\nu)=(25.47 \pm 
0.13) \% $ \cite{Davier05}. This discrepancy can be attributed mainly to 
the  fact that we have neglected the contribution of the $\rho(1450)$ 
resonance which affects the higher energy tail of the hadronic spectrum 
(indeed, if we repeat the fit to data of reference \cite{aleph} using the 
two vector resonance model in eq. 7, fixing the mass and width of the 
$\rho'$ 
to their PDG values \cite{pdg} and assuming that $\sigma$ is real, we get 
$m_{\rho}=775.80$ MeV, $g_{\rho\pi\pi}=5.867$ and $\sigma=-0.12$; this 
in turn leads to $B(\tau^- \to \pi^-\pi^0\nu_{\tau})=23.27\% $, 
which is closer to the experimental value). 

  It is worth to mention that in ref. \cite{Cirigliano02}, the authors 
have used the following form factor:
\be
f_+^{GP}(\tilde{t})=\frac{\sqrt{2}m_{\rho}^2}{m_{\rho}^2-\tilde{t}- 
im_{\rho}\Gamma^{GP}_{\rho}(\tilde{t})} 
\exp[2\tilde{H}_{\pi\pi}(\tilde{t}) 
+\tilde{H}_{K\bar{K}}(\tilde{t})]\ , 
\ee
which is obtained \cite{pichguerrero} by matching the prediction of 
chiral perturbation 
theory at $O(p^4)$ with the contribution of the $\rho(770)$ in the 
resonance region. The expressions of 
$\Gamma^{GP}_{\rho}(\tilde{t})$ 
(which differs from our decay rate given in section 2) 
and of the loops 
functions $\tilde{H}_{PP'}(\tilde{t})$ can be found in 
refs. 
\cite{Cirigliano02,pichguerrero}. As in our model, the form factor 
in eq. (11) gives a good 
description of  experimental data for the two-pion spectra in the region 
below $\sqrt{\tilde{t}} =1.1$ GeV.

The branching ratio for the non-radiative decay that is obtained using the 
form factor in eq. (11) also underestimates the experimental value since:
\be
B(\tau \to \pi\pi\nu)=21.19 \% \ .
\ee
This low branching ratio  reflects again the fact that the form factor 
in eq. (11) underestimates experimental data of the pion form factor for 
large values of $\tilde{t}$. One possibility to account for this 
discrepancy in the predictions of our model is to 
normalize our results for radiative decays in terms of  the 
non-radiative rate. However,  for the  purposes of  comparing 
our results with those of ref. \cite{Cirigliano02} we  keep the 
one-resonance model with the $\rho^-$ contribution in the evaluation 
of the model-independent radiative amplitudes.

\section{4. Radiative decay mode}

\begin{figure}
  \includegraphics[width=12cm]{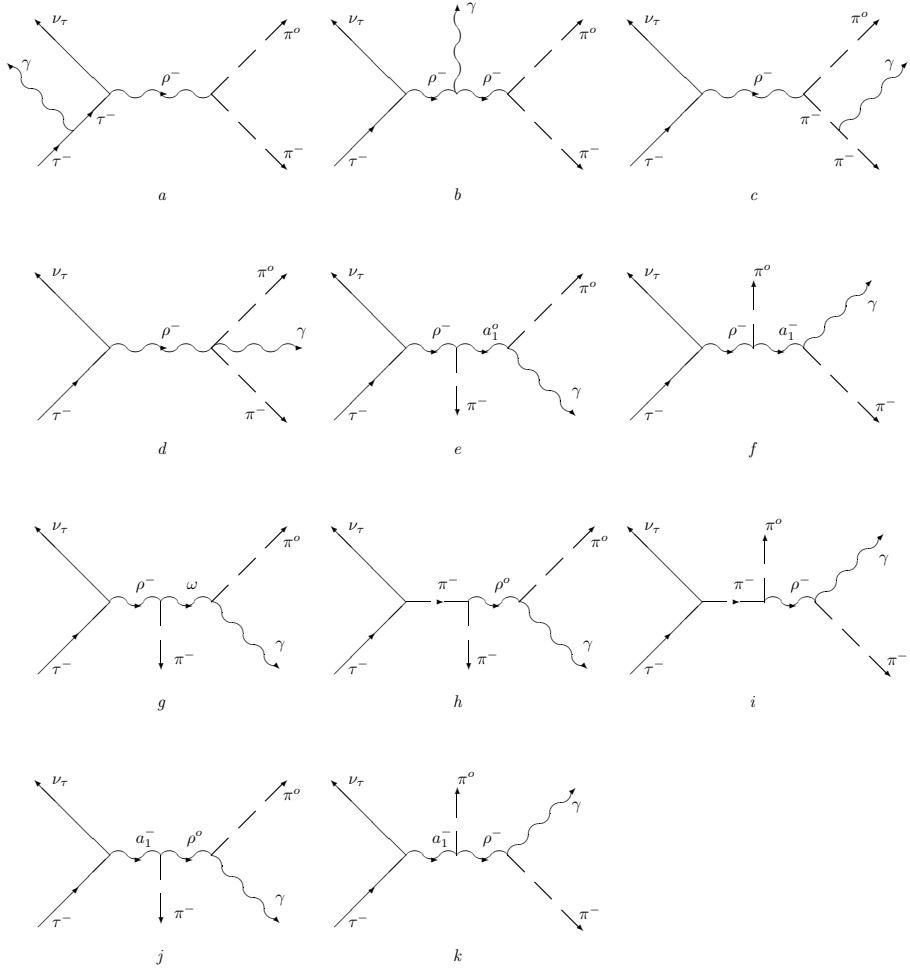}\\
  \caption{Feynman diagrams of the model-independent (a-d) and 
model-dependent (e-k) contributions to $\tau^- \to 
\pi^-\pi^0\nu_{\tau}\gamma$ decays.}\label{fig2}
\end{figure}

  The Feynman diagrams that contribute to the radiative $\tau^-(p) \to 
\pi^-(p_-)\pi^0(p_0)\nu_{\tau}(q)\gamma(k, \epsilon)$ decay in our 
meson dominance model are shown in Figure (\ref{fig2}). 
The particles four-momenta are indicated within parenthesis, with $k 
(\epsilon)$ denoting the momentum and polarization four-vectors of the 
photon. 

The  decay amplitude has the following generic expansion in powers of the 
photon energy $E_{\gamma}$ \cite{Low}: 
\be
{\cal M}=\frac{\cal A}{E_{\gamma}} + {\cal B}E_{\gamma}^0 + {\cal 
C}E_{\gamma} + \cdots, \
\ee
where the ellipsis denotes the terms of higher order in $E_{\gamma}$. 
As we will see below, the 
terms of order up to $E_{\gamma}^0$ (Low's amplitude) contains only 
model-independent 
contributions, while the terms starting at order $E_{\gamma}$ arise 
from model-dependent contributions and from the magnetic dipole 
($\beta(0)$)
and electric quadrupole moments ($Q_{\rho}$) of the 
$\rho^-$ meson (in this paper we do not consider the possible effects of 
$Q_{\rho}$).
In the following we consider the different contributions in more detail.

\subsection{A. Model-independent contributions}

The model-independent contributions (Fig. 2.(a-d)) are obtained by  
attaching the photon to all the charged  lines and vertices with 
derivate  couplings in the non-radiative Feynman diagram. This set of 
diagrams leads to a gauge-invariant amplitude. In our model, gauge 
invariance is guaranteed owing to the Ward identity satisfied by the  
electromagnetic coupling and propagator of the $\rho^-$ 
introduced in section 2. In other words, we 
do not need to {\it impose} gauge invariance to the 
model-independent amplitude  
due to the finite width effects of the $\rho^-$ vector meson. 
Just for a later comparison, let us mention that the effects of the 
$\rho^-$-meson magnetic moment $\beta(0)$, a 
gauge-invariant term by itself,  can not be obtained by imposing gauge 
invariance to the sum of amplitudes obtained from diagrams ($a, c, d$) in 
Figure 2.

\

Using the Feynman rules corresponding to the vertices and propagators in 
diagrams (a-d) from Figure 2, we obtain the following amplitudes:
\ba
{\cal M}_a &=&eG_FV_{ud} 
\frac{ig_{\rho}g_{\rho\pi\pi}}{\sqrt{2}}\frac{(p_--p_0)_\kappa  
D^{\beta\kappa}_{\rho^-}(p_-+p_0)}{2p\cdot k}\bar{u} 
(q)\gamma_{\beta}(1-\gamma_5) 
(\not p-\not k+m_{\tau})\not \epsilon^* u(p)  \\
{\cal M}_b&=& eG_FV_{ud} \frac{g_{\rho}g_{\rho\pi\pi}}{\sqrt{2}}
(p_--p_0)_{\nu}D^{\mu\nu}_{\rho^-}(p_-+p_0) 
\Gamma_{\kappa\eta\mu}D^{\delta\eta}_{\rho^-}
(p-q)\epsilon^{*\kappa} l_{\delta} \\
{\cal M}_c &=& 
-eG_FV_{ud}\frac{ig_{\rho}g_{\rho\pi\pi}}{\sqrt{2}} 
\frac{p_-\cdot \epsilon^*}{p_-\cdot k} 
(k+p_--p_0)_{\eta}D^{\delta\eta}_{\rho^-}(p-q)l_{\delta} \\
{\cal M}_d &=& 
eG_FV_{ud}\frac{ig_{\rho}g_{\rho\pi\pi}}{\sqrt{2}}\epsilon^*_{\eta} 
D^{\delta\eta}_{\rho^-}(p-q)l_{\delta} 
\ .
\ea

Owing to the Ward Identity given in section 2, it is easy to verify that 
the 
model-independent amplitude ${\cal M}_{MI}={\cal M}_a+{\cal M}_b+{\cal 
M}_c+{\cal M}_d$ is gauge-invariant, namely ${\cal M}_{MI}(\epsilon^* \to 
k)=0$. The amplitude ${\cal M}_{MI}$ differs from the corresponding 
model-independent amplitude of ref. \cite{Cirigliano02} in terms of order 
$k$ and due to the effects of the magnetic dipole  moment of the 
$\rho^-$ meson. As we will see later, the effects of $\beta(0)$ are  
negligible in the  integrated observables of this radiative decay. 
However, as it was discussed  elsewhere \cite{genaro00, nos}, its effects 
can be enhanced with an  special choice of the kinematics (see section 
5.D).

  Just to end this section, we provide the Low's amplitude obtained from 
Eqs. (14-17) after expanding the amplitude ${\cal M}_{MI}$ around the 
soft-photon limit (the form factor $f_+(t)$ used here corresponds to the 
expression in eq. (7) when $\sigma=0$):
\ba
{\cal M}_{Low} & = & \frac{e G_F V_{ud}}{\sqrt{2}} \bigg\{ f_+(t) \left( 
\frac{\epsilon^*
\cdot p_-}{k\cdot p_-} - \frac{\epsilon^* \cdot p}{k\cdot p}  \right)
(p_- - p_0)^\nu l_\nu + \frac{f_+(t)}{2k\cdot p} \bar{u}(q)(\not p_- -
\not p_0)\not k\not\epsilon^*(1-\gamma_5)u(p) \nonumber\\ & &  -f_+(t)
\left( \epsilon^{*\nu} - \frac{\epsilon^* \cdot p_-}{k\cdot p_-} k^\nu
\right) l_\nu + 2\frac{\mathrm{d} f_+(t)}{\mathrm{d}t} \left(
\frac{\epsilon^* \cdot p_-}{k\cdot p_-}k\cdot p_0 - \epsilon^* \cdot p_0
\right) (p_- - p_0)^\nu l_\nu \bigg\}  \ ,
\ea
As it can be easily checked, this amplitude coincides with the one 
obtained in Ref. \cite{Cirigliano02}. As is dictated by Low's soft-photon 
theorem \cite{Low}, the amplitude depends only on the non-radiative 
amplitude and on the static electromagnetic properties of the external 
particles.

\

\subsection{B. Model-dependent contributions}

  The model-dependent contributions that appear within our meson dominance 
model are shown in Figures 2.(e-k). The diagrams (e-g) contribute 
to an effective vector hadronic current, while the diagrams (h-k) give 
rise to 
an effective axial current. We can write these vector 
and axial model-dependent contributions to the amplitude as follows:
\ba
{\cal M}_V &=& eG_FV_{ud}\epsilon^{*\mu} \left[ 
V^e_{\mu\delta}+V^f_{\mu\delta}+V^g_{\mu\delta}\right] l^{\delta} \\
{\cal M}_A &=& eG_FV_{ud}\epsilon^{*\mu} \left[
A^h_{\mu\delta}+A^i_{\mu\delta}+A^j_{\mu\delta} +A^k_{\mu\delta}\right] 
l^{\delta} \ .
\ea 
The explicit expressions for the vector and axial terms of the 
hadronic vertex are the following:
\ba
V^e_{\mu\delta}&=& \frac{g_{\rho}g_{\rho a_1\pi}g_{\gamma
a_1\pi}}{\sqrt{2}e}\left[ {k\cdot
p_0} g_{\lambda\mu}-{k_\lambda p_{0\mu}} \right]
D_{a_1}^{\kappa\lambda}({k+p_0}) \left[ {(k+p_0)\cdot
(p-q)}g_{\eta\kappa}-{(k+p_0)_\eta (p-q)_\kappa} \right]
\nonumber\\ & & \times {D^{\,\,\,\quad\eta}_{\rho^- \delta}(p-q)} \\
V^f_{\mu\delta}&=&\frac{g_{\rho}g_{\rho a_1\pi}g_{\gamma
a_1\pi}}{\sqrt{2}e}\left[ {k\cdot
p_-} g_{\lambda\mu}-{k_\lambda p_{-\mu}} \right]
D_{a_1}^{\kappa\lambda}({k+p_-}) \left[ {(k+p_-)\cdot
 (p-q)} g_{\eta\kappa}-{(k+p_-)_\eta (p-q)_\kappa} \right] \nonumber\\
& & \times {D^{\,\,\,\quad\eta}_{\rho^- \delta}(p-q)} \\
V^g_{\mu\delta}&=&\frac{g_{\rho}g_{\rho \omega \pi}g_{\gamma
\omega\pi}}{\sqrt{2}e} 
\epsilon_{\lambda'\lambda\mu'\mu}{p_0}^{\lambda'}
{k}^{\mu'}D_{\omega}^{\kappa
 \lambda}({k+p_0})\epsilon_{\eta'\eta\kappa'\kappa}
{p_-}^{\eta'}({k+p_0})^{\kappa'}
{D^{\,\,\,\quad\eta}_{\rho^- \delta}(p-q)} \\ 
A^h_{\mu\delta}&=&
\frac{f_{\pi}g_{\rho\pi\pi}g_{\rho\pi\gamma}}{\sqrt{2}e{[(p-q)^2-m_\pi^2]}}
\epsilon_{\eta'\eta{\mu'}\mu}{p_0}^{\eta'}
 {k^{\mu'}}D_{\rho^-}^{{\kappa\eta}}({k+p_0})({p - q +
p_-})_{\kappa}({p-q})_\delta
\\
A^i_{\mu\delta}&=&
 \frac{f_{\pi}g_{\rho\pi\pi}g_{\rho\pi\gamma}}{\sqrt{2}e{[
(p-q)^2-m_\pi^2]}}
\epsilon_{\eta'\eta{\mu'}\mu}{p_-}^{\eta'}
{k^{\mu'}}D_{\rho^-}^{{\kappa\eta}}({k+p_-})({p-q+
p_0})_{\kappa}({p-q})_\delta
\\
A^j_{\mu\delta}&=&
\frac{if_{a_1}g_{\rho a_1\pi}g_{\rho\pi\gamma}}{\sqrt{2}e}
\epsilon_{\lambda'\lambda\mu'\mu}{p_0}^{\lambda'}{k}^{\mu'}
D_{\rho^-}^{\kappa\lambda}({k+p_0})  \left[ {(k+p_0)\cdot
(p-q)}g_{\eta\kappa}
-({k+p_0})_\eta ({p-q})_\kappa\right]
{D_{a_1\delta}^{\,\,\quad\eta}(p-q)} \\
A^k_{\mu\delta}&=&
\frac{if_{a_1}g_{\rho a_1\pi}g_{\rho\pi\gamma}}{\sqrt{2}e}
\epsilon_{\lambda'\lambda\mu'\mu}{p_-}^{\lambda'}{k}^{\mu'}
 D_{\rho^-}^{\kappa\lambda}({k+p_-}) \left[ {(k+p_-)\cdot (p-q)}
g_{\eta\kappa}
 -({k+p_-})_\eta ({p-q})_\kappa\right]
{D_{a_1\delta}^{\,\,\quad\eta}(p-q)} \ .
\ea

All the couplings constants appearing in the above expressions can be 
easily identified from the corresponding diagrams in Figure 2. Their 
values can be fixed from measured decays of the $a_1, \ \pi,\ \rho$ 
and  $\omega$ mesons and will be provided in the next section. Note that 
when the vector and axial vector mesons become heavy degrees of freedom, 
these model-dependent contributions vanish as required by chiral symmetry 
\cite{Cirigliano02}. 

  As  already anticipated, the above amplitudes are of order one in the 
photon four-momentum $k$. Moreover, they are individually 
gauge-invariant 
since the conditions $k^{\mu} 
V^m_{\mu\delta}=k^{\mu}A^n_{\mu\delta}=0$ are satisfied. Of 
course, the vector 
and axial amplitudes given above can be decomposed in terms of a basis of 
four-independent vector and axial tensors as pointed out in ref. 
\cite{Cirigliano02}. 

\section{5. Decay observables}

  As it was discussed in the previous section, the decay amplitude of the 
radiative $\tau$ lepton decay depends on a large set of parameters 
(coupling constants 
and masses of mesons). The parameters $m_{\rho^-}, \ g_{\rho\pi\pi}$ 
entering the 
model-independent contributions where fixed from a fit to experimental 
data \cite{aleph} in the di-pion mass  spectrum of the decay $\tau \to 
\pi\pi\nu$ in the region  below 1.1 GeV. Their values  were given in eq. 
(9) of section 3. The other free parameter entering the model-independent 
amplitude, namely  the  magnetic dipole moment $\beta(0)$ of the 
$\rho^-$ meson, is left as a free  parameter in order to study their 
effects on the different observables of radiative $\tau$ decays. 

   The model-dependent contributions depend mainly on the values of the 
coupling constants and masses of the vector ($\rho,\ \omega$), axial $a_1$ 
and pseudoscalar $\pi$ mesons (we assume isospin symmetry for the masses 
of neutral and charged states). The values of the coupling constants can 
be 
obtained from the measured decay rates of these mesons (except for the 
weak coupling of the $a_1$ meson whose value is fixed from the Weinberg 
sum rule $f_{a_1}=g_{\rho}$ \cite{okun}). As it 
was the case for the non-radiative decay in section 
3, we expect that these effective couplings will give a good estimate 
of the correct size for model-dependent effects. Based on the 
experimental data compiled  in ref. \cite{pdg}, we will use the following 
central values:

\ba
g_{\rho}&=& 167765.48\ \mbox{\rm MeV} \ \\  
f_{\pi} &=&  130.7\ \mbox{\rm MeV}\  \\
g_{\rho a_1\pi} &=& 4.843 \times 10^{-3}\ \mbox{\rm MeV}^{-1}\ \\\
g_{\gamma a_1\pi} &=& 2.9265 \times 10^{-4} \ \mbox{\rm MeV}^{-1}\\
g_{\rho\omega\pi} &=& 0.012\ \mbox{\rm MeV}^{-1}\\
g_{\gamma\omega\pi} &=& 7.1126 \times 10^{-4}\ \mbox{\rm MeV}^{-1}\\
g_{\rho\pi\gamma}&=& 2.2092\times 10^{-4} \ \mbox{\rm MeV}^{-1}\ .
\ea

Once we have fixed the values of these parameters, we proceed to compute 
the different observables associated to the radiative $\tau$ lepton decay. 
Using the 
choice of kinematical variables described in the Appendix, we can write 
the differential decay rate in terms of the five independent 
kinematical variables as 
follows:
\begin{equation}
\mathrm{d}\Gamma = \frac{\beta_{-0}}{2(4\pi)^6 m_{\tau}^2}
\frac{1}{2}\sum_{pols} |{\cal M}|^2 \mathrm{d}x
\mathrm{d}t \mathrm{d}E_{\gamma} \mathrm{d}\cos\theta_{\pi^-}
\mathrm{d}\phi_{\pi^-} \, ,
\end{equation}
where $\beta_{-0}=\sqrt{1-4m_{\pi}^2/t}$ is the magnitude of the pion 
velocity in the di-pion rest frame. In order to integrate over the 
relevant kinematical variables we have used the VEGAS 
\cite{vegas} integration routine. Next we  focus on the results obtained 
for each one of the computed observables.

\subsection{A. Branching ratios}

  In this subsection we compute the predictions of our model for 
the branching ratios. As it was done in ref. 
\cite{Cirigliano02}, we distinguish between the {\it bremsstrahlung} 
(model-independent) and the {\it full} 
(that includes also model-dependent terms) contributions to the decay 
rate.  Since the 
unpolarized probability is divergent for soft photons, we introduce a 
cutoff energy $E_{\gamma}^{min}$ to regularize the integral. In Figure 3 
we  show the branching ratio as a function of $E_{\gamma}^{min}$, for the 
normal value of the magnetic dipole moment $\beta(0)=2$. We compare our 
results  with  the predictions based on chiral perturbation 
theory \cite{Cirigliano02} (the three squares in Figure 3). 

 We observe that if we exclude the $\omega$ meson contribution, 
diagram in Figure 2.g, we find a very good agreement with the calculation 
of reference \cite{Cirigliano02}. However, 
according to the VMD model, 
the contribution of the $\omega(782)$ vector meson can not be excluded. 
This particular model-dependent contribution becomes large due 
to a particular kinematic accident, namely the almost degeneracy of the 
$\rho^- -\omega$ system, and due to the small decay width of the $\omega$ 
meson ($\Gamma_{\omega}=8.44$ MeV). This double resonance effect produces 
an enhancement of the 
decay amplitude in approximately the same kinematical region. In order 
to verify this explanation we have increased the mass difference of the 
$\rho^-\omega$ mesons and/or the width of the $\omega$ meson and have 
found that the large effect of the $\omega$ meson is decreased in an 
important way. We find 
that for 
photon cutoff  energies of order $E^{min}_{\gamma} =200$ MeV, the 
contribution 
of the $\omega$ meson 
becomes already twice the value of all other contributions. Therefore, a 
measurement of the radiative decay branching ratio can help to discriminate 
among the two models.

\begin{figure}
  \rotatebox{270}{\includegraphics[width=10cm]{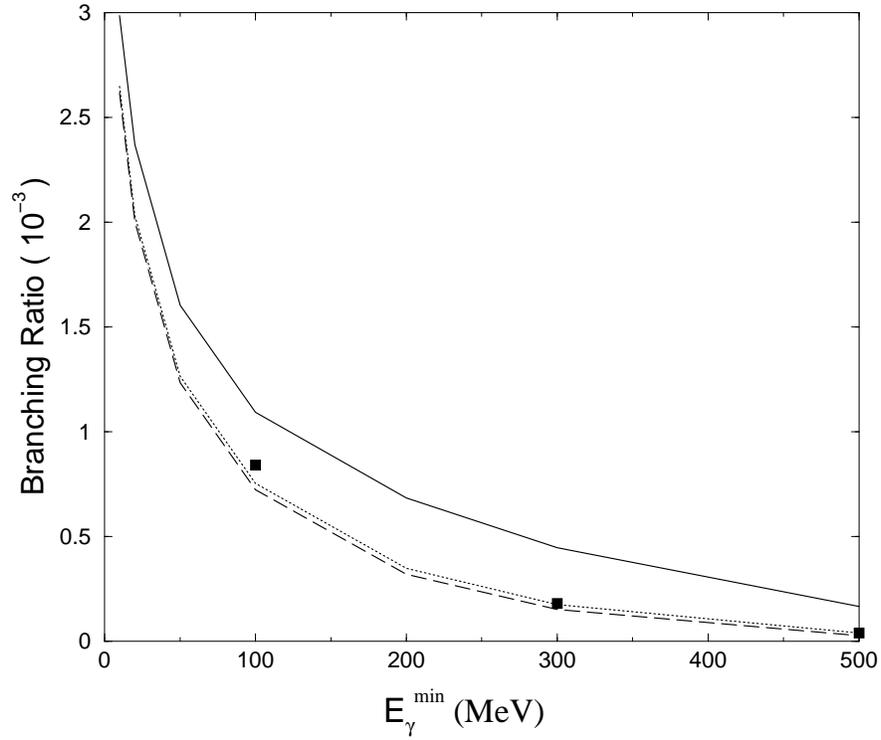}}\\
  \caption{Branching ratio of $\tau \to 
\pi^-\pi^0\nu\gamma$ as a function of the soft-photon cutoff 
$E_{\gamma}^{min}$ for $\beta(0)=2$. The dashed-line denotes 
the model-independent contributions and the 
solid-line the full contributions. The dotted-line is obtained by 
excluding the contribution of  
the $\omega$ meson, diagram 2.g.  The points at $E_{\gamma}= 
100,\ 200$ and $300$ MeV correspond to the 
full contributions of ref. \cite{Cirigliano02}.}\label{fig3} 
\end{figure}

   The branching ratio is almost insensitive to reasonable variations in 
the value of the $\rho^-$-meson magnetic dipole moment $\beta(0)$. In 
Figure 4 we show the full branching ratio as a function of 
$E^{min}_{\gamma}$ for  $\beta(0)= 1, 2$ and $3$. Thus, it is clear that 
this 
observable cannot  help to discriminate values of the magnetic dipole 
moment.

\begin{figure}
  \rotatebox{270}{\includegraphics[width=10cm]{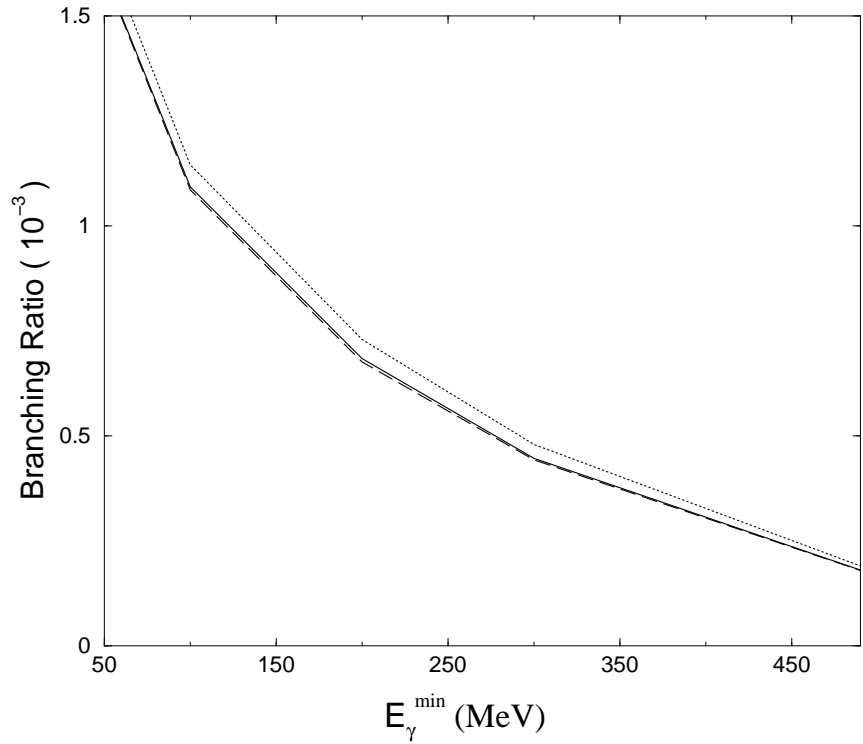}}\\
  \caption{Branching ratio of $\tau \to \pi^-\pi^0\nu\gamma$ as a 
function of the soft-photon cutoff  $E_{\gamma}^{min}$ for 
$\beta(0)=1,2,3$ (respectively, dashed, solid and dotted lines). Only the 
full contributions are plotted in this case. }\label{fig4} 
\end{figure}

\subsection{B. Photon spectrum}

  The photon spectrum can be obtained after integrating over all the 
kinematical variables in eq. (35) except $E_{\gamma}$. This spectrum 
$d\Gamma/dE_{\gamma}$ is plotted 
in Figure 5 for $\beta(0)=2$. The effect of the $\omega$ meson 
contribution 
is particularly important for $E_{\gamma} \geq 180$ MeV. As in the case of 
the 
branching ratio, the photon spectrum is not sensitive to  the value of the 
$\rho^-$ magnetic dipole moment.  

\begin{figure}
  \rotatebox{270}{\includegraphics[width=10cm]{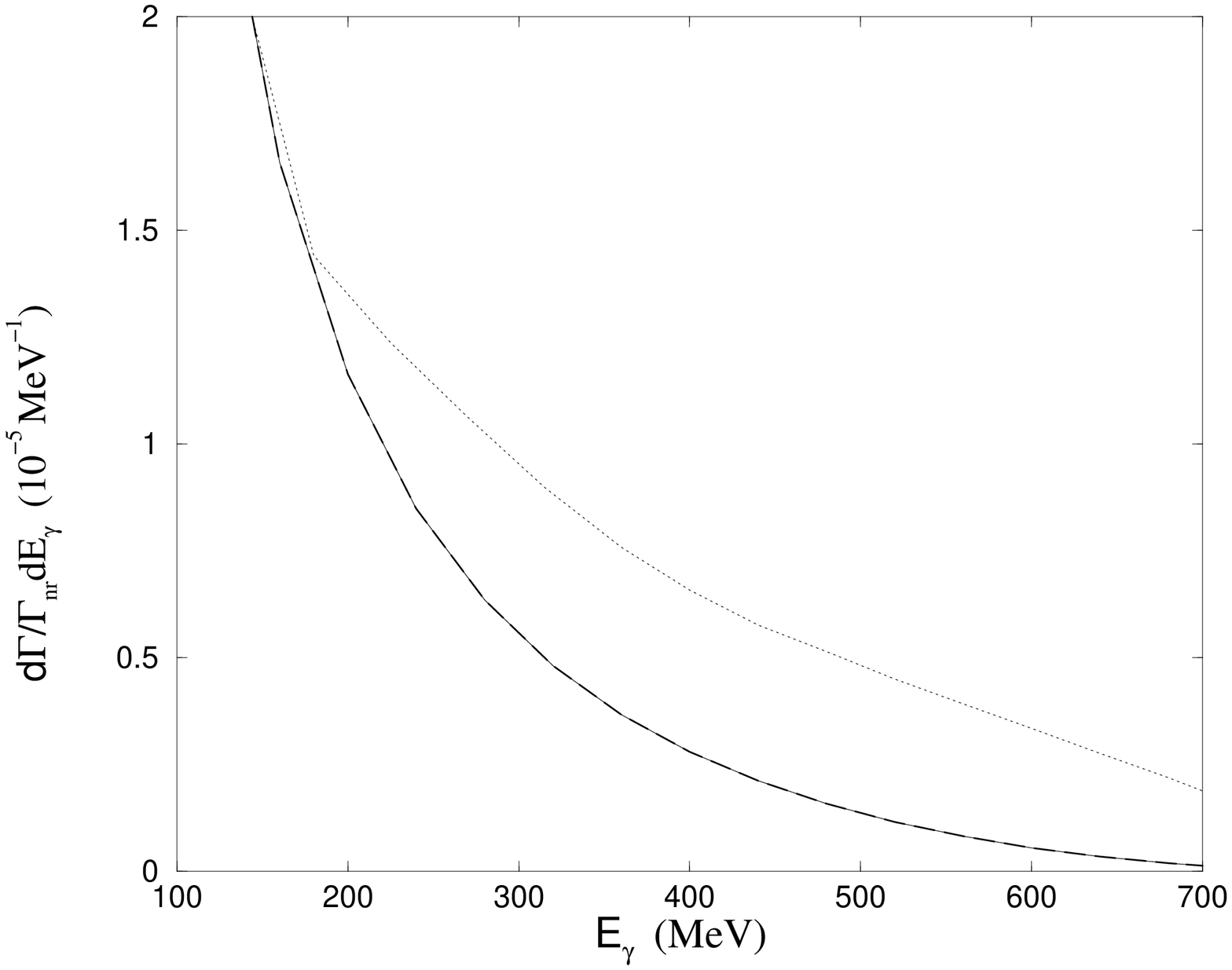}}\\
  \caption{Photon spectrum in the $\tau \to \pi^-\pi^0\nu\gamma$ decay. 
The solid-line denotes the model-independent contributions, while the 
dotted-line is used for the full contributions. The dashed-line 
(almost superposed over the solid-line) corresponds to the full 
contribution obtained by excluding fig. 2.g. The observable 
is normalized to the non-radiative decay rate.}\label{fig5} 
\end{figure}

\subsection{C. Di-pion invariant mass distribution}

  Another important observable associated to $\tau^- \to 
\pi^-\pi^0\nu\gamma$ is the invariant mass distribution of the di-pion 
system. In the case of the corresponding non-radiative decay, this 
spectrum shows explicitly the peaks associated to the production 
of vector resonances. It is 
interesting to study how they are modified by the radiation of 
photons. On another hand, a detailed study of this spectrum in radiative 
decays is very important in order to remove the hard bremsstrahlung from 
photon inclusive measurements of $\tau^- \to \pi^-\pi^0\nu (\gamma)$ 
decays \cite{Davier05}.

   In Figure 6 we plot the combined photon and di-pion invariant 
mass spectra $d\Gamma/\Gamma_{nr}dE_{\gamma}dt$  ($\Gamma_{nr}$ is the 
non-radiative decay rate) by choosing $\beta(0)=2$. In order 
to avoid the infrared divergences due to the emission of soft-photons, we 
plot our results for a finite value of the  photon energy $E_{\gamma}$. 
Once 
again, we observe that the presence of the 
$\omega$-meson contribution changes the spectrum in a sizable way in 
all the region of $t$. However, the position of the peaks associated to 
the photon emission off the $\pi^-$ and the $\tau^-$ external lines are 
not affected.  We also  plot in Figure 7 the di-pion invariant mass 
distribution after integrating the previous result for 
photons of energy larger than 300 MeV (unfortunately, a 
quantitative comparison with results of reference \cite{Cirigliano02} is 
not possible since they give their results in arbitrary units). As in the 
case of the branching 
ratios and  of the photon spectrum studied in previous subsections, the 
di-pion invariant mass 
distribution can also help to distinguish between the present model and 
the 
one based in 
chiral perturbation theory \cite{Cirigliano02}.

\begin{figure}
  \rotatebox{270}{\includegraphics[width=10cm]{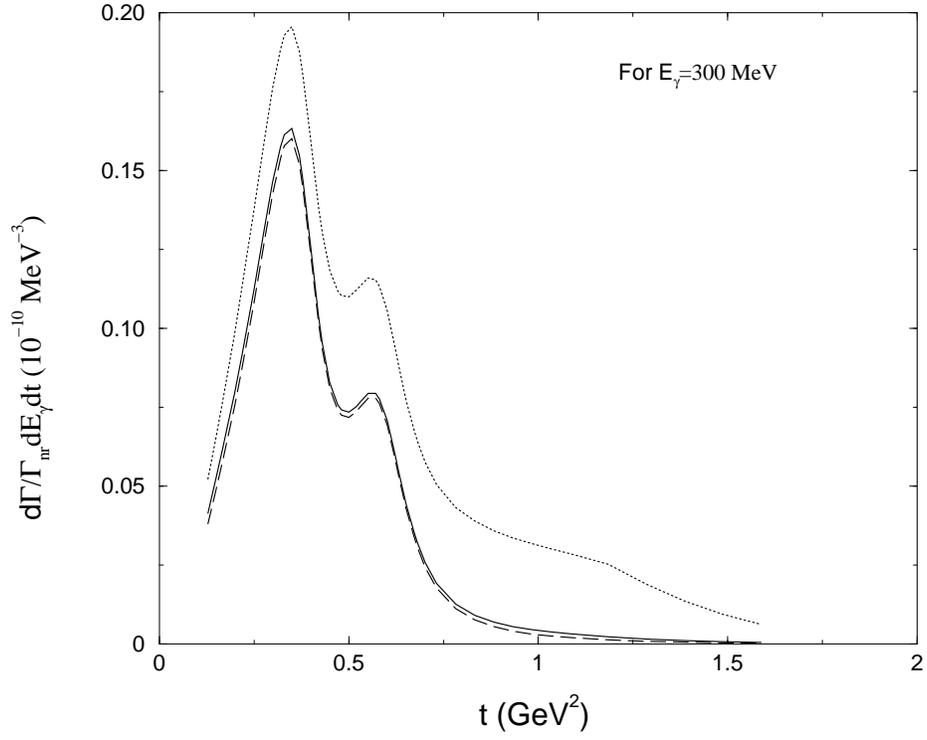}}\\
  \caption{Distribution in the photon energy and the invariant mass of the 
di-pion system in $\tau \to \pi^-\pi^0\nu\gamma$ decays for 
$E_{\gamma}=300$ MeV. The dashed-line 
(dotted-line) denotes the model-independent (full) contributions. The 
solid-line is obtained when we exclude the meson $\omega$, 
diagram in fig. 2.g. The observable is normalized to the 
non-radiative rate.}\label{fig6} 
\end{figure}

\begin{figure}
  \rotatebox{270}{\includegraphics[width=10cm]{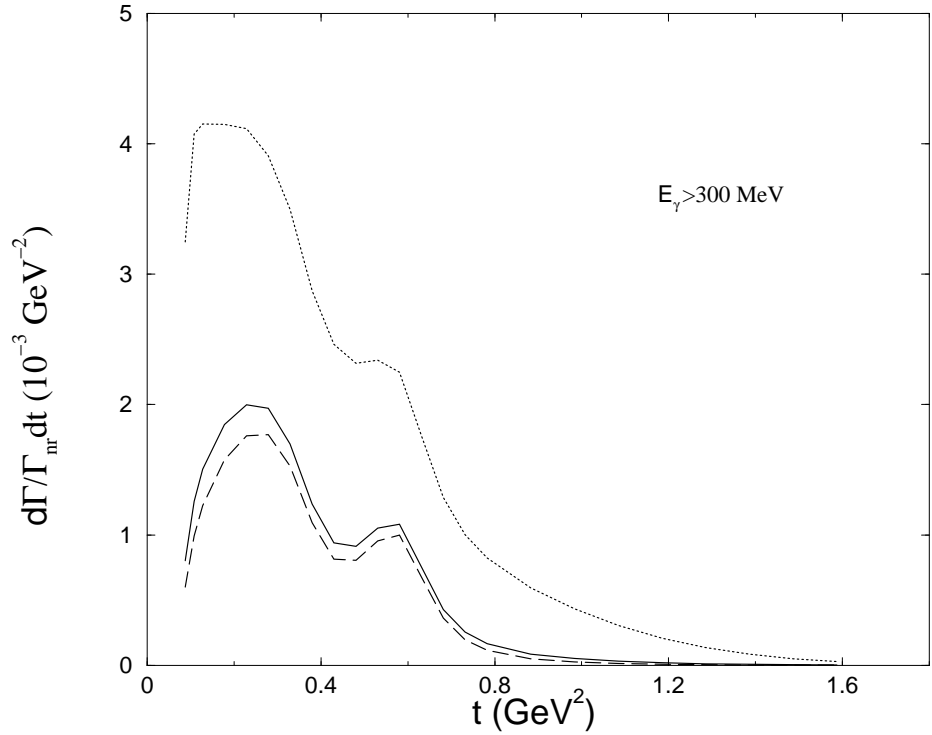}}\\
  \caption{Di-pion invariant mass distribution in  $\tau \to 
\pi^-\pi^0\nu\gamma$ decays for photon energies larger than  
300 MeV. Description of lines are the same as in Figure 
6. The observable is normalized to the non-radiative rate.}\label{fig7} 
\end{figure}

\subsection{D. Angular and energy photon spectra}

Previous studies of radiative decays involving the production and 
decay of and on-shell charged vector meson 
\cite{nos} have shown that the angular and energy photon spectra are 
sensitive to the effects of the vector-meson magnetic dipole moment 
when photons are emitted at small angles. 
Therefore, we also compute this observable for the case of $\tau 
\to \pi\pi\nu\gamma$ decays.

   In Figure 8 we plot the angular and energy photon spectra 
$d\Gamma/\Gamma_{nr}d\cos \theta dE_{\gamma}$ as a function of the photon 
energy for 
$\theta=10^0$ and $20^0$ ($\theta$ is the angle between the photon and 
the $\pi^-$ in the $\tau^-$ lepton rest frame) and three different values 
of $\beta(0)$. We have subtracted the 
(well known) contribution arising from the pure bremsstrahlung (terms of 
order $k^{-2}$ in the unpolarized probability) in order to make more 
visible the  effect of $\beta(0)$. We observe that there are 
some kinematical regions where the sensitivity to $\beta(0)$ is increased 
and it may eventually help to measure this property. 

\begin{figure}
  \rotatebox{270}{\includegraphics[width=10cm]{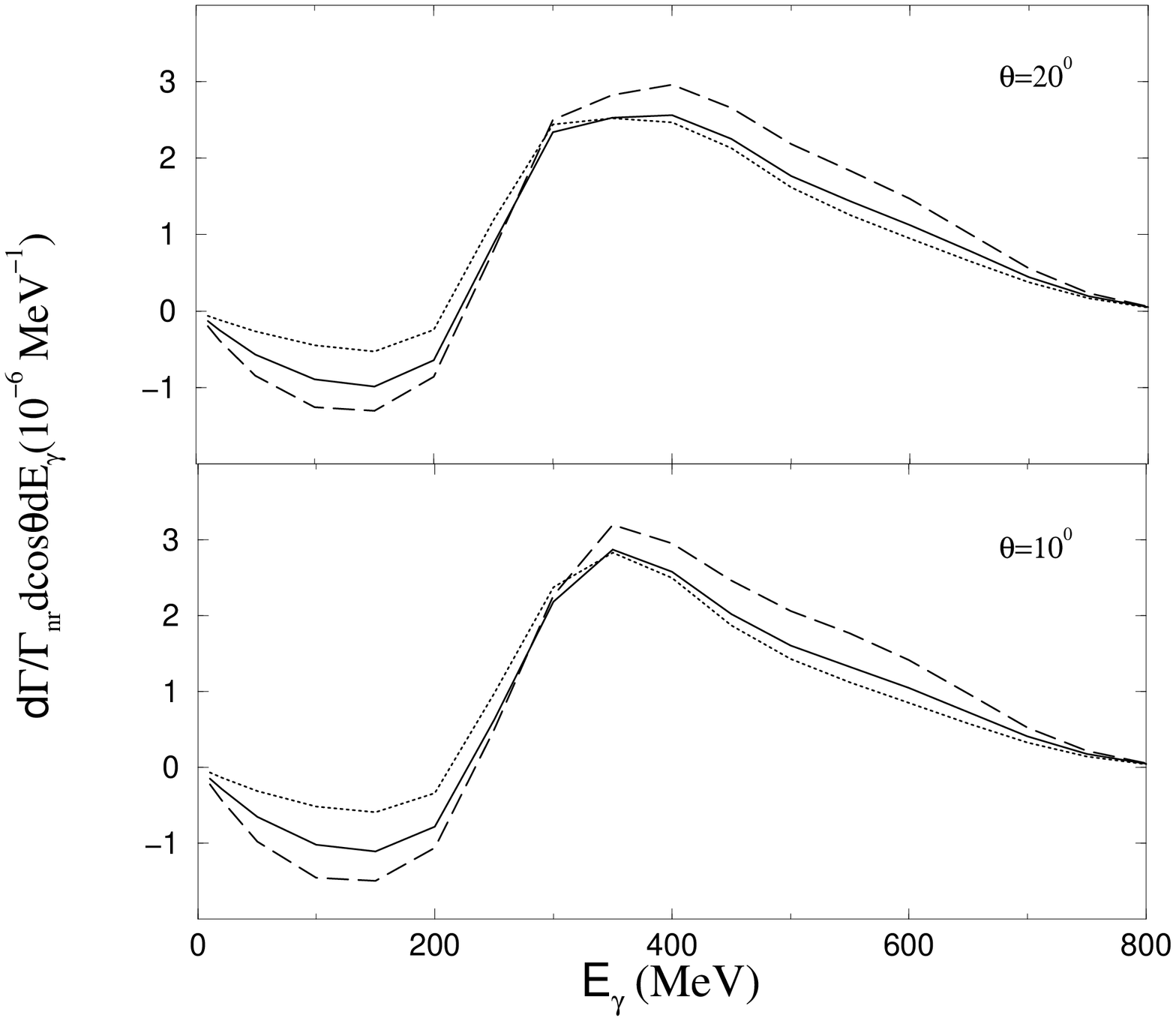}}\\
  \caption{Reduced angular and energy distribution of photons for 
three different values of $\beta(0)=1,2,3$ (dotted, solid and 
dashed lines) and two different angles of the photon emitted with respect 
to the $\pi^-$ three-momentum. The model-independent contributions (for 
$\beta(0)=0$) have been subtracted. The observable is 
normalized to the non-radiative rate.}\label{fig8} 
\end{figure}

\section{6. Conclusions}

   In this paper we have considered the radiative two pion decay of the  
$\tau$ lepton.  This decay mode was considered previously in refs. 
\cite{genaro00, Cirigliano02}. The new ingredients of the present study 
include, $(a)$ an electromagnetic gauge-invariant description of 
the model-independent 
amplitude including intermediate unstable $\rho^-$ mesons and, $(b)$ a 
complete 
calculation of the model-dependent contributions using a meson dominance 
model.  
In the framework of the present model, all the meson states that 
are allowed to contribute as intermediate particles were included
in the calculation of the radiative amplitude. In addition, we study
the effects of the $\rho^-$ magnetic dipole moment in the observables 
of the radiative $\tau$ lepton decay.

  We have found that the branching ratio, the photon spectrum and the
di-pion invariant mass  spectrum of radiative $\tau$ lepton decays are
sensitive to the model-dependent contributions that include an $\omega$ meson
intermediate state (Figure 2.g). This contribution produces an important enhancement of
these observables with respect to all other contributions 
arising in the present model. The origin of this enhancement can be traced 
back to the almost
degeneracy of the $\rho-\omega$ masses and due to the small decay width of
the $\omega$ meson. In the absence of this contribution, our model 
reproduces the results obtained in the framework of chiral perturbation 
theory \cite{Cirigliano02}. Since
present formulations of the chiral lagrangian interactions do not include
the presence of vector-vector-pseudoscalar vertices, it is 
natural that the calculation of reference \cite{Cirigliano02} 
has not included the contribution of diagram 2.g. Thus, experimental 
measurements of the observables studied in this paper can help to assess 
the approximation involved in different models. On another hand, our 
calculation confirms that the effects of the model-dependent axial 
contributions are negligible \cite{Cirigliano02}.

  The quantitative difference in model-dependent terms may 
modify the size of the corrections applied to extract the pion form factor 
from photon inclusive measurements of  $\tau \to \pi\pi\nu\gamma$ decays. 
As is  well known \cite{Davier05}, this pion form factor seems to be 
a bit larger than the one obtained from $e^+e^- \to \pi^+\pi^-$ 
annihilations for squared momentum transfers larger than $m_{\rho}^2$, 
even after known sources of isospin breaking corrections 
are taken into account.
Since this study must consider the effects of virtual radiative 
corrections, we will consider it elsewhere \cite{next}. 

  The different observables studied in this paper are not sensitive to 
the effects of the $\rho^-$ magnetic dipole moment. As it was pointed out 
in ref. \cite{genaro00} 
the photon emission off the internal charged $\rho$ meson line (Figure 
2.b) is expected
to carry information about this important property which has not been 
measured yet. The sensitivity on different values of $\beta(0)$ is 
slightly increased when we consider the angular and energy photon spectra 
for almost collinear $\pi^- -\gamma$ particles\footnote{ As it was concluded 
in refs. \cite{nos}, choosing such small angles help to suppress 
radiation off electric charges and make more prominent the radiation off the 
magnetic dipole moment.}. Thus, only processes where the charged $\rho$ 
vector-mesons 
are on their mass-shell can offer a better sensitivity to the 
magnetic dipole moment \cite{nos}, since the radiation off this 
electromagnetic moment enters at a lower order in the photon momentum.

\section{Acknowledgments}
   The authors acknowledge financial support from Conacyt under the grants 
40473, 41600 
and 42026. G.T. Acknowledges the grant PAPIIT IN112902-3 from UNAM.

\

\section{Appendix: Kinematics}

  We discuss here the kinematics of the decay $\tau^-(p)\to 
\pi^-(p_-)\pi^0(p_0)\nu(q)\gamma(k)$; for simplicity we choose the 
isospin limit $p_0^2=p_-^2=m_{\pi}^2$.  The unpolarized squared amplitude 
for a four-body decay depends upon five 
independent kinematical variables. We choose this set of independent 
variables to be (we closely follow ref. \cite{pais}):
\be
(x, t, E_{\gamma}, \cos \theta_{\pi^-}, \phi_{\pi^-})\ .
\ee
The quantity $t= (p_-+p_0)^2\ (x=(q+k)^2)$ denotes the squared invariant 
mass of the two-pion ($\nu\gamma$) system, $E_{\gamma}$ is the 
photon-energy in  the 
rest frame of the $\tau$, and $(\theta_{\pi^-}, \phi^-)$ are  
the spherical coordinate angles that define the $\pi^-$ three-momentum in 
the $\tau$ lepton rest frame.

   The order of the limits of integration can be conveniently chosen 
according on the 
energy or angular distribution that we want to obtain for the observables. 
We consider four  possible choices (after integrating upon the angular 
variables):
\begin{itemize}
\item If we integrate successively on $E_{\gamma},\ t$ and $x$, 
the  limits of integration are given by:
\ba
m_{\gamma}^2 \leq &x& \leq (m_{\tau}-2m_{\pi})^ 2 \\
4m_{\pi}^2 \leq &t& \leq (m_{\tau}-\sqrt{x})^2 \\
\frac{m_{\tau}^2+x-t-2X}{4m_{\tau}}  \leq &E_{\gamma}& \leq                    
\frac{m_{\tau}^2+x-t+2X}{4m_{\tau}}
\ea
where $2X=\sqrt{\lambda(m_{\tau}^2, x, t)}$, and $m_{\gamma}$ 
is  a cutoff parameter introduced to regularize the infrared divergence.
\item If we exchange $x \leftrightarrow t$ in the order of 
integration of the previous case, their corresponding limits are:
\ba
4m_{\pi}^2 \leq &t& \leq (m_{\tau}-m_{\gamma})^2 \\
m_{\gamma}^2 \leq &x& \leq (m_{\tau}-\sqrt{t})^2\ .
\ea
\item The successive order of integration over $t,\ x$ and 
$E_{\gamma}$, 
requires  the integration region to be defined as:
\ba
m_{\gamma} \leq &E_{\gamma}& \leq \frac{m_{\tau}^2-4m_{\pi}^2}{2m_{\tau}} 
\\
0 \leq &x& \leq 
\frac{2E_{\gamma}(m_{\tau}^2-4m_{\pi}^2-2m_{\tau}E_{\gamma})}{m_{\tau}- 
2E_{\tau}} 
\\ 4m_{\tau}^2 \leq &t& \leq 
\frac{(m_{\tau}-2E_{\tau})(2m_{\tau}E_{\gamma}-x)}{2E_{\gamma}}\ .
\ea
\item Another useful choice is (the order of integration is 
easily understood):
\ba
m_{\gamma} \leq &E_{\gamma}& \leq \frac{m_{\tau}^2-4m_{\pi}^2}{2m_{\tau}} 
\\
4m_{\pi}^2 \leq &t& \leq m_{\tau}(m_{\tau}-2E_{\gamma}) \\
0\leq &x& \leq \frac{2E_{\gamma}(m_{\tau}^2-2m_{\tau}E_{\gamma} 
-t)}{m_{\tau}-2E_{\gamma}} \ .
\ea
\end{itemize}
Other choices for the order of integrations are also possible. To  
verify that the different domains of integrations are equivalent, we have 
performed the numerical integrations using the different domains of 
integration described above and have verified that the same results are 
obtained.

\end{document}